\magnification=\magstep1
\baselineskip=16pt
\hfuzz=6pt

$ $

\vskip 1in

\centerline{\bf Quantum approximate optimization is computationally universal}

\bigskip

\centerline{Seth Lloyd}

\centerline{Departments of Mechanical Engineering and Physics}

\centerline{Massachusetts Institute of Technology} 

\centerline{Xanadu}

\bigskip
\noindent{\it Abstract:} The quantum approximate optimization algorithm (QAOA)
applies two Hamiltonians to a quantum system
in alternation.  The original goal of the algorithm 
was to drive the system close to the ground state of one of the 
Hamiltonians.    This paper
shows that the same alternating procedure can be used to perform universal
quantum computation: the times for which the Hamiltonians are applied
can be programmed to give a computationally universal dynamics.   
The Hamiltonians required can be as simple as homogeneous sums of single-qubit
Pauli $X$'s and two-local $ZZ$ 
Hamiltonians on a one-dimensional line of qubits.

\vskip 1cm

Quantum information processing supports a broad range of platforms
ranging from universal quantum computers capable of 
performing quantum algorithms such as factoring, to quantum annealers,
which are not computationally universal but which can be used
to try to find answers to hard optimization problems.   The quantum
approximate optimization algorithm (QAOA) is a dynamic
optimization method that is related to quantum annealing [1-2].   
As originally proposed, QAOA is not obviously computationally universal. 
This paper shows that QAOA is capable of universal quantum computation
in a simple and natural way.

The quantum approximate optimization algorithm operates by
applying two different Hamiltonians to a quantum system in alternation to try
to drive the system to the ground state of one of the Hamiltonians [1-2].
In its original form,
the first Hamiltonian $H_Z = {\rm poly}({Z_j})$ is a low-order
polynomial in the Pauli $Z$ operator over the qubits, and
the second Hamiltonian $H_X = \sum_j X_j$ is a uniform sum
of Pauli $X$ operators.   Starting in the uniform superposition
of logical qubits, $|I\rangle = 2^{-n/2} \sum_{j=0}^{2^n-1} |j\rangle$,
one first applies $H_Z$ for time $t_1$, then $H_X$ for time $\tau_1$,
then $H_Z$ for time $t_2$, then $H_X$ for time $\tau_2$, and so on,
$p$ times in alteration, yielding the state  
$$ U(\vec t, \vec \tau) |I\rangle
=  e^{-i\tau_p H_X} e^{-it_p H_Z} \ldots  e^{-i\tau_2 H_X} e^{-it_2 H_Z}
e^{-i\tau_1 H_X} e^{-it_1 H_Z} |I\rangle.\eqno(1)$$
Because of its alternating form, the procedure for performing the QAOA
is sometimes referred to as the quantum alternating operator ansatz.

The original purpose of QAOA was to 
vary the $t$'s and $\tau$'s for fixed $p$ to try to make
$U(\vec t, \vec \tau) |I\rangle$
approximate the ground state of $H_Z$.   This procedure works
rather well, even for small $p$ [1-2].   The form
of the QAOA dynamics (1) exhibits a variety of features.  The
alternating form of the application of operators in QAOA 
makes it an application of `bang-bang' quantum control [3-4],
which like its classical cousin, is known to be time-optimal via
the Pontryagin minimum principle [5].   Its simplicity and
flexibility means that
QAOA can be repurposed for
investigations of quantum supremacy/advantage [6], and for
quantum search [7].    

In this paper I show that QAOA represents a natural framework
for performing universal quantum computation.   In particular,
when $H_Z$ is a simple, homogeneous two-qubit Hamiltonian
on a one-dimensional lattice, the times $\vec t$, $\vec \tau$
can be selected to program the system to implement any
desired sequence of quantum logic gates.   The method is
to apply QAOA to implement computationlly universal
broadcast quantum cellular automaton architectures [8-11]. 

The Hamiltonians required to implement universal quantum computation
via QAOA are particularly simple.  Let
$H_X = \sum_j X_j$ as before, and let
$$\eqalign{
H_Z =& \sum_j \omega_A Z_{2j} + \omega_B Z_{2j+1} +
\gamma_{AB} Z_{2j} Z_{2j+1} + \gamma_{BA} Z_{2j+1} Z_{2j+2}\cr
\equiv & ~ \omega H_A + \omega_B H_B + 
\gamma_{AB} H_{AB} +  \gamma_{BA} H_{BA},\cr}
\eqno(2)$$
where $\omega_A, \omega_B, \gamma_{AB}, \gamma_{BA}$ are
not rationally related.   We can then choose $t$ to effectively
`turn on' one of the Hamiltonians $H_A$, $H_B$, $H_{AB}$,
or $H_{BA}$, while `turning off' the others.      For example,
choose $t$ so that
$$|\omega_A t -( 2\pi n_A + \phi) < \epsilon/4, 
 |\omega_B t - 2\pi n_B| < \epsilon/4, |\gamma_{AB} t - 
2\pi n_{AB}| < \epsilon/4, |\gamma_{BA} t -      
2\pi n_{BA}| < \epsilon/4, \eqno(3)$$ 
where $n_A$, $n_B$, $n_{AB}$, $n_{BA}$ are integers.
The amount of time it takes to attain this accuracy
is $O( 1/\epsilon^4 )$: with four incommensurate
Hamiltonians one must `wrap around' $O(\epsilon^4)$ times
to line up the appropriate phases to accuracy $\epsilon$.
The resulting transformation obeys 
$$ \| e^{-it H_Z} - e^{-i\phi_A H_A} \|_1 <  \epsilon.\eqno(4)$$ 
In a similar fashion, we can implement the transformations
$e^{-i\phi_B H_B}$, $e^{-i\phi_{AB} H_{AB}}$, $e^{-i\phi_{BA} H_{BA}}$
to any desired degree of accuracy.   

Implementing  $e^{-i\phi_A H_A}$, $e^{-i\phi_B H_B}$,
and $e^{-i\phi_{AB} H_{AB}}$ transformations allows us to construct
transformations of the form
$$ U_{AB} = e^{-i\phi_A H_A} e^{-i\phi_B H_B}e^{-i\phi_{AB} H_{AB}}
= U_{01} \otimes U_{23} \otimes \ldots \otimes U_{2j,2j+1} \otimes
\ldots \eqno(5) $$
where $U_{2j, 2j+1}$ can be any desired two-qubit unitary with determinant
equal to 1, diagonal in the $Z$ basis,  
that acts on pairs of spins $2j, 2j+1$.
Similarly, we can construct transformations of the form
$$ U_{BA}= e^{-i\phi_A H_A} e^{-i\phi_B H_B}e^{-i\phi_{BA} H_{BA}}
= U_{12} \otimes U_{34} \otimes \ldots \otimes U_{2j+1,2j+2}\otimes 
\ldots \eqno(6) $$
where again $U_{2j+1, 2j+2}$ can be any desired unitary
diagonal in the $Z$ basis.

By adjusting the $\tau$ terms in the QAOA formula (1), we
see that we can also implement transformations of the form
$$e^{-i3\pi X/4} e^{-it H_Z} e^{-5\pi X/4} =
e^{-i t H_Y}, \eqno(7)$$
where $H_Y$ has the same form as $H_Z$ in equation (2), but
all the $Z$ Pauli matrices have been transformed into $Y$ Pauli matrices.
Consequently, we can implement transformations of the form
(5) and (6), but where the unitaries $U_{2j, 2j+1}$,
$U_{2j+1, 2j+2}$ are now diagonal in the $Y$ basis.
But the ability to perform transformations of the form (5)
and (6) with the $U$'s diagonal in $Z$ or in $Y$ basis
implies that one can perform transformations of the form
(5) and (6) for {\it any} desired $U_{2j, 2j+1}$,
$U_{2j+1, 2j+2}$. 

To summarize, the QAOA procedure of equation (1) allows
us to implement a {\it broadcast quantum cellular automaton} dynamics
[8-11] of the form
$$U^p_{BA} U^p_{AB} \ldots U^1_{BA} U^1_{AB},\eqno(8)$$
where the $U^k_{AB}$, $U^k_{BA}$ can be varied at will
from step to step.     If, in addition to being able
to apply the broadcast quantum CA dynamics of equation (8),
one can measure and prepare the first qubit of the one-dimensional
array, one can perform universal quantum computation using
well-established methods developed in [8-11].
These methods operate by first implementing pulse sequences
that load data onto the array, next, by applying pulses
that implement a parallel quantum computation, and finally
by moving the results of the computation to the
first qubit of the array where they can be measured
out sequentially.

\bigskip\noindent{\it Discussion:}

This paper showed that the dynamics of 
quantum approximate optimization algorithm can be programmed
to perform any desired quantum computation.
Note that the results of the paper imply that the even
simpler dynamics are quantum computationally universal.
In particular, since the Hadamard operation transforms the
$Z$ Hamiltonian into the corresponding operator with $Z$ Pauli matrices
transformed into $X$'s, the programmable dynamics
 $$U(\vec t) 
=  H e^{-it_p H_Z} \ldots H  e^{-it_2 H_Z}
 H e^{-it_1 H_Z}, \eqno(9)$$
where $H= H_1 \otimes \ldots \otimes H_j \ldots$ is
the tensor product of single qubit Hadamard operations,
is also computationally universal.   Equation (9) 
represents the repeated application of an instantaneous quantum polynomial
time dynamics (IQP), using the same $Z$ Hamiltonian each time [12].
Our results show that such IQP dynamics with fixed $Z$ Hamiltonian
is computationally universal.  

The architecture used to prove universality here was a simple
one-dimensional array with nearest neighbor interactions: 
such one-dimensional architectures do not support thresholds
for scalable fault-tolerant quantum computation.    The
Hamiltonian averaging trick introduced here can easily
be extended to higher-dimensional arrays, and long-range
interactions could be added to implement long-range
`wires' to transmit quantum information between distant
parts of the network.    Such architectures might be
able to support fault-tolerant quantum computation via QAOA. 
For each additional type of interaction
Hamiltonian added, the time required for each `active' parallel quantum logic
operation is multiplied by an additional factor of $1/\epsilon$ so
that all the `non-active' interactions can wrap around.   
How many different types of wires/interactions are required
to give a network with a desired degree of long-range connectivity is
an open question.

Although the QAOA dynamics is phrased in terms of turning
on and turning off interactions, a physical implementation
of the computationally universal dynamics exhibited here
could be implemented with an always-on $Z$ Hamiltonian,
with the $X$ rotations or Hadamards implemented by a single,
strong, globally applied pulse.    The simplicity of the
addressing required to perform universal quantum computation
in such systems suggests their implementation via superconducting
systems, arrays of atoms in optical lattices, spin, or quantum
dot systems.      Even when such systems are not capable of
scalable quantum computation, their ability to exhibit quantum
supremacy/advantage suggests that they could be used to construct
near term quantum information processing devices for problems
such as deep quantum learning [13], where the weights of the deep
quantum network are given by the adjustable times $\vec t$.
For example, the goal of the deep quantum learning procedure
could be to train the computationally universal QAOA system
to implement a unitary transformation that maps inputs
to outputs, given a training set of such input-output pairs.
Or the system could be trained to try to create quantum states
on which measurement results match the statistics of a classical
data set.

\vfill 
\noindent{\it Acknowledgements:} This work was supported by 
IARPA under the QEO project.
\vfil\eject

\noindent{\it References:}

\bigskip\noindent [1] E. Farhi, J. Goldstone, S. Gutmann,
`A quantum approximate optimization algorithm,'
arXiv: 1411.4028. 

\bigskip\noindent [2] E. Farhi, J. Goldstone, S. Gutmann, 
`A Quantum Approximate Optimization Algorithm Applied to a 
Bounded Occurrence Constraint Problem,' arXiv: 1412.6062. 

\bigskip\noindent [3] L. Viola, S. Lloyd,
`Dynamical suppression of decoherence in two-state quantum systems,'
{\it Phys. Rev. A} {\bf 58}, 2733 (1998);
arXiv: quant-ph/9803057.  

\bigskip\noindent [4] L. Viola, E. Knill, S. Lloyd,
`Dynamical decoupling of open quantum systems,' 
{\it Phys. Rev. Lett.} {\bf 82}, 2417 (1999);
arXiv: quanti/ph/9809071.

\bigskip\noindent [5] E. Farhi, A.W. Harrow,
`Quantum Supremacy through the Quantum Approximate Optimization Algorithm,'
arXiv: 1602.07674.

\bigskip\noindent [6] Z-C. Yang, A. Rahmani, A. Shabani, H. Neven, 
C. Chamon,  `Optimizing variational quantum algorithms using 
Pontryagin's minimum principle,' arXiv: 1607.06473.

\bigskip\noindent [7] 
Z. Jiang, E. Rieffel, Z. Wang, `A QAOA-inspired circuit for 
Grover’s unstructured search using a transverse field,' 
arXiv: 1702.02577.

\bigskip\noindent [8]
S. Lloyd, `A potentially realizable quantum computer,'
{\it Science} {\bf 261}, 1569-1571 (1993).

\bigskip\noindent [9]
S. Lloyd, `Programming pulse driven quantum computers,'
arXiv: quant-ph/9912086. 

\bigskip\noindent [10]
S. Benjamin, `Schemes for parallel quantum computation without local control of qubits,'
{\it Phys. Rev. A} {\bf 61}, 020301 (2000); arXiv: quant-ph/9909007.

\bigskip\noindent [11]
S. Benjamin, `Quantum Computing Without Local Control of 
Qubit-Qubit Interactions,'
{\it Phys. Rev. Lett.} {\bf 88}, 017904 (2001).

\bigskip\noindent [12]
M.J. Bremner, R. Jozsa, D.J. Shepherd, `Classical simulation of 
commuting quantum computations implies collapse of the polynomial hierarchy,'
{\it Proc. Roy. Soc. A} {\bf 467}, 459–472 (2010); arXiv: 1005.1407.

\bigskip\noindent [13] J. Biamonte, P. Witteck, N. Pancotti, P. Rebentrost,
N. Weibe, S. Lloyd, `Quantum machine learning,' {\it Nature} {\bf 549},
195-202 (2017).

\vfill\eject\end